\documentclass[structabstract,letter]{aa}  
\usepackage{graphicx}
\usepackage{txfonts}
\usepackage{natbib}
\bibpunct{(}{)}{;}{a}{}{,} 
\usepackage{color}
\usepackage{paralist}

\newcommand{\mch}{\ensuremath{\mathrm{M}_\mathrm{Ch}}}
\newcommand{\msun}{\ensuremath{\mathrm{M}_\odot}}
\newcommand{\nuc}[2]{\ensuremath{^{#1}\mathrm{#2}}}

\newcommand{\ye}{\ensuremath{Y_\mathrm{e}}}
\newcommand{\gcc}{\ \mathrm{g\ cm^{-3}}}

\begin{document}
\title{Solar abundance of manganese: a case for the existence of near Chandrasekhar-mass Type Ia supernova progenitors}


\author{Ivo~R.~Seitenzahl$^{1,2}$, Gabriele~Cescutti$^{3}$, Friedrich~K.~R\"opke$^{1}$, Ashley~J.~Ruiter$^{2}$,
R\"udiger~Pakmor$^{4}$}

\institute{Institut f\"ur Theoretische Physik und Astrophysik,
  Universit\"at W\"urzburg, Campus Hubland Nord,
  Emil-Fischer-Str. 31,\\
  D-97074 W\"urzburg, Germany\\
  \email{iseitenzahl@astro.uni-wuerzburg.de}\and 
  Max-Planck-Institut f\"ur Astrophysik,
  Karl-Schwarzschild-Str. 1, D-85741 Garching, Germany\and
Leibniz-Institut f\"ur Astrophysik Potsdam (AIP), An der Sternwarte 16, D-14482 Potsdam,
  Germany\and 
  Heidelberger Institut f\"{u}r Theoretische Studien,
  Schloss-Wolfsbrunnenweg 35, D-69118 Heidelberg, Germany }

\date{Received xxxx xx, xxxx / accepted xxxx xx, xxxx}


\abstract
{Manganese is predominantly synthesised in Type Ia supernova (SN~Ia)
  explosions.  Owing to the entropy dependence of the Mn yield in
  explosive thermonuclear burning, SNe~Ia involving near Chandrasekhar-mass (\mch) white
  dwarfs (WDs) are predicted to produce Mn to Fe ratios significantly exceeding 
  those of SN~Ia explosions involving sub-Chandrasekhar mass primary WDs.
  Of all current supernova explosion models, only SN~Ia models involving near \mch\
  WDs produce [Mn/Fe] $\gtrsim 0.0$.
}
{Using the specific yields for competing SN~Ia
  scenarios, we aim to constrain the relative fractions of exploding near-\mch\
to sub-\mch\ primary WDs in the Galaxy. }
{We extract the Mn yields from three-dimensional thermonuclear
supernova simulations referring to different initial setups and progenitor channels.
 We then compute the chemical evolution of Mn in the Solar neighborhood,
  assuming SNe~Ia are made up of different relative fractions of the
  considered explosion models.}
{We find that due to the entropy dependence of freeze-out yields from nuclear statistical equilibrium,
[Mn/Fe] strongly depends on the mass of the exploding WD, with near-\mch\ WDs producing substantially 
higher [Mn/Fe] than sub-\mch\ WDs.
Of all nucleosynthetic sources potentially influencing the chemical evolution of Mn, only
explosion models involving the thermonuclear incineration of near-\mch\ WDs predict solar
or super-solar [Mn/Fe]. 
Consequently, we find in our chemical evolution calculations
 that the observed [Mn/Fe] in the Solar neighborhood at [Fe/H]
  $\gtrsim 0.0$ cannot be reproduced without near-\mch\ SN~Ia primaries.
  Assuming that $50\%$ of all SNe~Ia stem from explosive thermonuclear 
  burning in near-\mch\ WDs results in a good match to data. }
{}


\titlerunning{Solar Mn abundance requires near-\mch\ SN~Ia progenitors}
\authorrunning{Seitenzahl et al. 2013} 
\maketitle
%

\section{Introduction}
There is a general consensus that thermonuclear explosions of carbon-oxygen
WDs are the underlying physical process leading to Type Ia supernova (SN~Ia)
explosions \citep[for a recent review on SNe~Ia
see, e.g.,][]{hillebrandt2013a}.  In spite of this general agreement on the
basic underlying physical picture, neither the exact explosion
mechanism(s) nor the formation channel(s) of binary stellar evolution
leading up to the explosion have reached a consensus model.  Loosely
speaking, two main evolutionary scenarios have emerged. In the \textit{single degenerate} scenario (SDS) first described by
\citet{whelan1973a}, a WD accretes mass from a stellar companion until
it explodes following the onset of a carbon fusion runaway as it
approaches the Chandrasekhar-mass (\mch) limit.  Recent multi-dimensional
simulations of explosions of near-\mch\ WDs
include pure deflagration
\citep[e.g.][]{roepke2007a,jordan2012b,ma2013a,fink2013a}, deflagration-to-detonation
transition,
\citep[e.g.][]{gamezo2005a,roepke2007b,bravo2008a,kasen2009a,seitenzahl2011a,seitenzahl2013a}
pulsational reverse detonation \citep[e.g.][]{bravo2009a}, and
variants of gravitational confined detonation models
\citep[e.g.][]{plewa2007a,meakin2009a,jordan2012a}.
In the \textit{double degenerate} scenario (DDS) first proposed by
\citet{iben1984a} and \citet{webbink1984a}, the progenitor system is
a binary system of two WDs.  For sufficiently close
binaries, the emission of gravitational waves will lead to orbital 
decay, potentially resulting in a thermonuclear explosion triggered by the merger of the two WDs.
Proposed explosion mechanisms in the DDS can be divided into two categories, depending on the
existence of an accretion torus.
(1) Although it is generally believed that accretion from the thick disc around the primary
 \citep[e.g.][]{tutukov1979a,mochkovich1990a} leads to its collapse  
to a neutron star \citep[e.g.][]{nomoto1991a,dessart2006b,yoon2007a} following 
its transformation to an O-Ne-Mg core \citep{saio1985a,timmes1994a,saio1998a},
\citet{piersanti2003b,piersanti2003a} and \citet{saio2004a} argue that for rapidly-rotating primaries, 
central carbon ignition may be possible. The latter case would result in a near-\mch\ SN~Ia event,
with the same potential explosion mechanisms listed above.
(2) Recent multi-dimensional hydrodynamical simulations have shown that an accretion disc need not
form and the resulting \emph{violent merger} of the two WDs may lead to a detonation in
the primary \citep{pakmor2010a,pakmor2011b,pakmor2012a,dan2011a,raskin2012a}.
In this violent merger model, the explosion is essentially driven by 
a pure detonation of a nearly-hydrostatic sub-\mch\ WD.

From the point of view of explosion modeling, the important
question is whether the primary WD is near-\mch\ (resulting from the SDS or mergers with accretion from a torus) or significantly sub-\mch\
(from violent mergers or double-detonations in He-accreting systems,
e.g. \citealt[][]{woosley1994b}).
\citet{mazzali2007a} argue for the former case, while
\citet{stritzinger2006a} support the latter.
We show that the two possibilities lead to significant
differences in the Mn to Fe production ratio, and we 
argue that a significant fraction of
Galactic SNe~Ia must arise from explosions of near-\mch\ WDs. 
We continue by analyzing the impact of the difference in
Mn on chemical evolution models and comparing the results to observational data on
Mn abundances in the Sun and in Galactic stars.

\section{Nucleosynthesis of M\lowercase{n} in SN~I\lowercase{a}}
\label{sec:nucleo}
A key focus of this work is on the production of manganese in
explosive nucleosynthesis. Mn (atomic number 25) has only one stable isotope,
\nuc{55}{Mn}.  Most of the \nuc{55}{Mn} produced in thermonuclear
explosive burning is synthesized as \nuc{55}{Co}
\citep[e.g.][]{truran1967a}, which then decays via \nuc{55}{Fe} to the
stable \nuc{55}{Mn}.
The two main nucleosynthetic processes synthesizing \nuc{55}{Co}, and
hence Mn, are ``normal'' freeze-out from nuclear statistical
equilibrium (NSE) and incomplete Si-burning. For freeze-out from NSE
to be ``normal'' as opposed to ``alpha-rich'', the mass fraction of
$\nuc{4}{He}$ has to remain rather low during the freeze-out phase
\citep[$\lesssim$1 per cent according to][]{woosley1973a}. For explosive
nuclear burning this is the case at relatively high density
\citep[$\rho \gtrsim 2 \times 10^8 \gcc$, see][]{thielemann1986a,
  bravo2012a}, which implies relatively low entropy. At lower density,
the \nuc{55}{Co} present in NSE is readily destroyed during the
alpha-rich freeze-out via $\nuc{55}{Co}(p,\gamma)\nuc{56}{Ni}$
\citep[see][]{jordan2003a}, resulting in a much lower final [Mn/Fe].
We note that a recent study has shown that the \nuc{55}{Co} to \nuc{56}{Ni}
production ratio is rather insensitive to nuclear reaction rate
uncertainties \citep{parikh2013a}.

To put this critical density into context, note that the mass of a cold WD
(\ye=0.5) in hydrostatic equilibrium with central density $\rho_c = 2
\times 10^8 \gcc$ is $M = 1.22\, \msun$.  Only explosions of near-\mch\
 WDs involve densities high enough to result in
``normal'' freeze-out from NSE. Violent mergers
\citep[][]{pakmor2012a} as well as sub-\mch\
double-detonations \citep[e.g.][]{fink2010a,kromer2010a} of typical SN~Ia
brighness have primary core masses below $1.2\, \msun$
\citep{sim2010a,ruiter2011a}.
We therefore have a robust, physical
reason for the large difference in [Mn/Fe]. 
Delayed-detonation models, which 
undergo significant thermonuclear explosive burning at densities above $\rho \gtrsim 2
\times 10^8 \gcc$ will have an enhanced production of Mn from
the contribution of ``normal'' freeze-out from NSE, which is not
the case for violent merger or double-detonation models. 
This division between ``normal'' and ``alpha-rich'' freeze-out
is also the reason for the predicted differences of the late-time bolometric light curves
\citep[][]{seitenzahl2009d,roepke2012a}.

We note that for very neutron-rich environments,
\nuc{55}{Mn} could also be directly synthesized. Therefore, it is
natural to ask the question of whether
gravitational settling of \nuc{22}{Ne} in sub-\mch\ WDs
can significantly affect our main point that [Mn/Fe] for SNe~Ia
resulting from these objects is significantly sub-solar. In contrast to canonical ignition in
near-\mch\ WDs, convective burning is not expeced to precede
the explosion here.  The potential effects of
concentrating neutron rich material near the WD's core are therefore
in principle possible. 
For gravitational settling to play a role
i) the sub-\mch\ WD has to remain liquid and 
ii) sufficient time must pass to allow for appreciable \nuc{22}{Ne} to fall 
from low density to high density regions where iron-group nucleosynthesis occurs.
That the sub-\mch\ primary WD in a DDS system 
remains liquid for the \nuc{22}{Ne} to settle is already unlikely,
since for cooling and non-accreting WDs the \nuc{22}{Ne}
settling time-scale ($t_s$) is longer than the crystallization time-scale in the core \citep{bildsten2001a}.
Even if the WD were to remain liquid, the relevant time-scales are too long to significantly affect our conclusions.
For example, for a hot ($T=10^8\,\mathrm{K}$) $1.2\,\msun$ WD, $t_s \approx 5\,\mathrm{Gyr}$, and for a
cold ($T=10^6\,\mathrm{K}$) $1.2\,\msun$ WD, $t_s \approx 23\,\mathrm{Gyr}$ \citep{bravo1992a}.
Furthermore, the settling time-scale $t_s$ is increasing strongly with decreasing WD mass \citep[e.g.][]{bildsten2001a}. Consequently, less massive WDs around $1.0\,\msun$ would show even less of an effect.
Since most SNe~Ia have much smaller delay times \citep[e.g.][]{maoz2012b},
we expect that gravitational settling of \nuc{22}{Ne} will not change our conclusions.

\section{Galactic chemical evolution of M\lowercase{n}}
\label{sec:gce}
Observational data
show that halo stars have an average abundance ratio for [Mn/Fe]$\,\sim\,
$$-0.5$ \citep[see][]{sobeck2006a},
providing a strong indication that SNe~II produce a sub-solar ratio of
Mn to Fe.  Theoretical nucleosynthesis calculations of massive stars agree with
these observational findings; most of the models
\citep[e.g.][]{woosley1995a,limongi2003a,nomoto2006a} predict
[Mn/Fe] yields typically three times lower than the one observed in the Sun.
The solar value for the mass ratio of Fe to Mn can be computed
from the photospheric abundances \citep{grevesse2010a} by assuming the
same mean atomic weights observed on Earth. Assuming
uncorrelated errors, we obtain for the elemental mass ratio $\mathrm{Fe}/\mathrm{Mn} = 119 \pm 15$.

SNe~Ia enrich the interstellar medium with a time delay compared to the first 
core-collapse SNe, which means that they did not significantly affect the
chemical evolution in the solar vicinity until $[\mathrm{Fe}/\mathrm{H}] \sim -1.0$
\citep[see e.g.][]{matteucci1986a}. Indeed, from around this metallicity,
[Mn/Fe] derived from observed stellar abundances displays a strong
increase \citep[e.g.][]{gratton1988a,gratton1991a}.
Although \citet{feltzing2007a} invoke strongly metallicity dependent SNe~II
Mn yields, the rise of [Mn/Fe] for $[\mathrm{Fe}/\mathrm{H}] \gtrsim -1.0$ 
 to the value observed in the Sun is typically 
attributed to the nucleosynthesis contribution of SNe~Ia \citep[e.g.][]{gratton1989a,timmes1995a,francois2004a,cescutti2008a,kobayashi2006a,kobayashi2009a,kobayashi2011b}.

We perform chemical evolution calculations (see Sec.~\ref{sec:results})
that only differ in the yields assumed for SNe~Ia (see Sec.~\ref{sec:yields}).
Our model for the solar vicinity, which is essentially the same as adopted in \citet{cescutti2008a},
is based on the model introduced by \citet{chiappini1997a} (called ``two infall model'').
We use for all cases the same delay time distribution (DTD) \citep{greggio1983a},
 although we are aware that this is a simplistic approach.
Assuming a different DTD for e.g. the merger scenario from analytical formalisms
\citep[e.g.\ as][]{greggio2005a} or binary evolution calculations
\citep{ruiter2009a} could modify the trend obtained by our chemical
evolution model. Examples of the sensitivity on the DTD can be found in
\citet{matteucci2009a} for the case of [O/Fe] and in
\citet{kobayashi2009a}.  However,
assuming yields for SNe~Ia lower
than solar will always result in a Mn to Fe ratio below the solar
value, independent of the assumed DTD.   
For the contribution of massive star explosions we
assume the metallicity dependent yields calculated by \citet{woosley1995a}.
We note that these yields do not substantially differ from the yields calculated by
other groups \citep[see e.g.][]{limongi2003a,nomoto2006a,kobayashi2011b}.
We did not include the contribution of low and intermediate mass stars here \citep[e.g.][]{pignatari2013a},
since they do not produce/destroy enough Mn or Fe to significantly affect our results.

\subsection{SN~Ia yield data}
\label{sec:yields}
We use different yields for near-\mch\ and sub-\mch\ explosion models.
As our main representative for near-\mch\ primaries (often likened to the SDS),
 we use the N100 model of a delayed detonation from
\citet{seitenzahl2013a}. For sub-\mch\ primaries,
we use the violent merger model of two WDs with 1.1 and 0.9\,\msun\ published in \citet{pakmor2012a},
which can also be thought of as a representative of the DDS. 
We have chosen these two models since they
produce rather typical \nuc{56}{Ni} masses of ${\sim}0.6\,\msun$ and 
have already been compared in their optical \citep{roepke2012a} and gamma-ray \citep{summa2013a} emission.
Due to a significant difference in central density,
the production of Mn is a factor of ${\sim}3$ smaller for
the merger-model compared to the delayed-detonation model (see Sec.~\ref{sec:nucleo} and Table~\ref{tab:1}).

\citet{pakmor2013a} suggest that all SNe~Ia derive from mergers of
two WDs, except for pure deflagrations in near-\mch\ WDs that leave bound remnants
behind -- a model that matches the observables of SN~2002cx-like SNe well 
\citep[see][]{phillips2007a,kromer2013a}.
We therefore also include the N5def model of \citet{fink2013a}. 

\section{Results}
\label{sec:results}
\begin{table}
  \caption{[Mn/Fe] yields for select thermonuclear (Ia), core collapse (II), and hypernova (HN) models of solar metallicity progenitors. Only models of near-\mch\ SNe~Ia predict $[\mathrm{Mn}/\mathrm{Fe}] \geq 0.0$.}
  \begin{tabular}{|l|l|l|c|c|} \hline model name & SN type &
    masses & [Mn/Fe] & ref. \\ \hline \hline
    N100          & Ia    & near-\mch\ &  0.33 & (1) \\
    N5def         & Ia    & near-\mch\ &  0.36 & (2) \\
    N150def       & Ia    & near-\mch\ &  0.42 & (2) \\
   W7            & Ia    & near-\mch\ &  0.15 & (3) \\
   W7            & Ia    & near-\mch\ &  0.02 & (4) \\
    \hline
    1.1\_0.9      & Ia    & sub-\mch\ & -0.15\tablefootmark{a} & (5) \\ 
    1.06 \msun    & Ia    & sub-\mch\ & -0.13\tablefootmark{a} & (6) \\
    \hline
    WW95B\tablefootmark{b}       & II    & $11 < M/\msun\ < 40$ & -0.15\tablefootmark{c}  & (7) \\
    LC03D\tablefootmark{d}        & II &  $ 13 < M/\msun\ < 35 $      & -0.27\tablefootmark{c}            & (8) \\ 
    N06        & II+HN &  $ 13 < M/\msun\ < 40 $     & -0.31\tablefootmark{c}            & (9) \\
    \hline
  \end{tabular}\\
\tablefoottext{a}{The given reference is for the explosion model; the respective [Mn/Fe]
 yields are published here for the first time, assuming that the main sequence 
 progenitor had a solar metallicity \citep{asplund2009a} and primary C, N, O was 
converted to \nuc{22}{Ne} during core He-burning.}\\
\tablefoottext{b}{We use model B for $M \geq 30\,\msun$.}
\tablefoottext{c}{Weighted with a Salpeter IMF.}
\tablefoottext{b}{We use model sequence D throughout.}

  \tablebib{(1)~\citet{seitenzahl2013a}; (2)~\citet{fink2013a};
    (3)~\citet{iwamoto1999a}; (4)~\citet{maeda2010a};
    (5)~\citet{pakmor2012a};
    (6)~\citet{ruiter2013a}; (7)~\citet{woosley1995a} ; (8)~\citet{limongi2003a} ; (9)~\citet{nomoto2006a}.} 
  \label{tab:1}
\end{table}
In Table~\ref{tab:1}, we have compiled a selection of [Mn/Fe] yields for 
different supernova types from the literature.
It is evident that currently only models involving thermonuclear explosions
of near-\mch\ WDs predict [Mn/Fe]$\,>\,$$0.0$.
Assuming that we are not missing a significant 
nucleosynthetic production site of Mn, this alone already tells us that near-\mch\ WDs primaries must 
contribute significantly to the production of Mn and Fe, and therefore constitute a significant fraction of
SNe~Ia. To corroborate this result and to place further constraints on the relative fractions of near-\mch\ and sub-\mch\
WD primaries, we consider five different chemical evolution cases, 
each case only differing in the nucleosynthetic yields assumed for SN~Ia as follows:

\begin{compactitem}
\item \emph{case $M_{Ch}$}: SN~Ia yields are from the N100 model of a delayed
  detonation in a near-\mch\ WD \citep{seitenzahl2013a}.
\item \emph{case $subM_{Ch}$}: SN~Ia yields are from the violent merger of a 1.1 with a 0.9\,\msun\ 
 WD \citep{pakmor2012a}.
\item \emph{case mix}: 50\% of SNe~Ia explode as in \emph{case $M_{Ch}$}
and 50\% as in \emph{case $subM_{Ch}$}. 
\item \emph{case $M_{Ch}+$}: similar to \emph{case $M_{Ch}$}, but SN~Ia yields
depend on progenitor metallicity \citep[using models $\mathrm{N100\_Z0.01}$, $\mathrm{N100\_Z0.1}$ and N100 from][]{seitenzahl2013a}.
\item \emph{case $subM_{Ch}$+2002cx}: 20\% of SNe~Ia explode as 
pure deflagrations leaving remnants \citep[model N5def from][]{kromer2013a}
 and the remaining 80\% explode as in \emph{case $subM_{Ch}$}.
\end{compactitem}

\begin{figure}
  \begin{center}
     \includegraphics[width=0.46\textwidth]{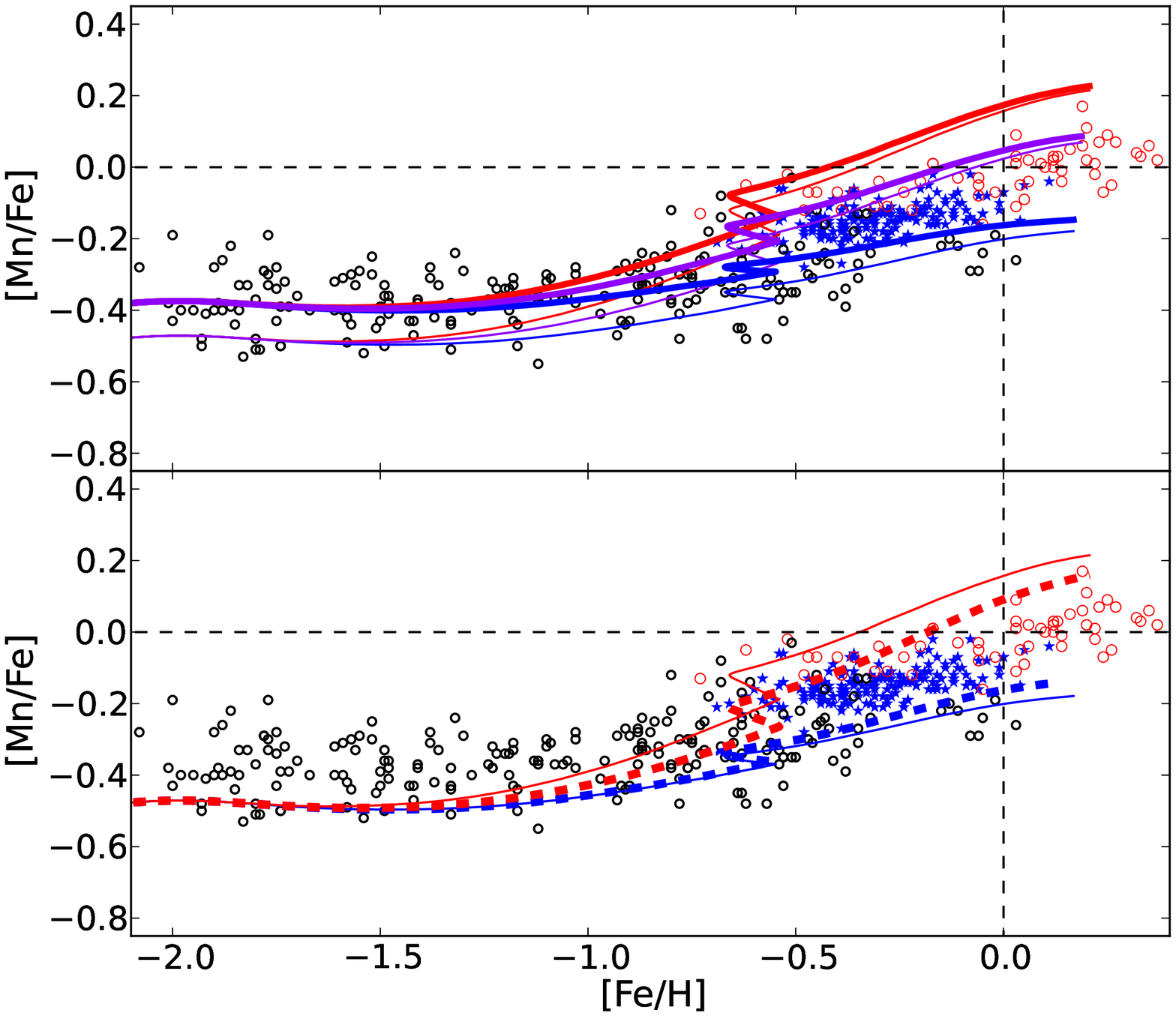}
    \caption{[Mn/Fe] vs [Fe/H] in the solar vicinity. 
      Open black squares are data from \citet{sobeck2006a},
       blue stars are from \citet{reddy2003a}, and
      red open dots are thin disc data from \citet{feltzing2007a}.
Top~panel: Thin lines are for massive star yields from
\citet{woosley1995a}, thick lines enhanced their Mn yields by 25 per cent.
Red lines are for \emph{case $M_{Ch}$}, blue lines for
      \emph{case $subM_{Ch}$}, and \emph{case mix} are the purple lines.
        Bottom~panel:
       Dashed thick blue line is
      for \emph{case $subM_{Ch}$+2002cx}, dashed thick red line is for
      \emph{case $M_{Ch}+$}. Thin blue and red lines are as in the top panel.
       }\label{fig1}
  \end{center}
\end{figure}

In Fig.\ \ref{fig1} (top), 
we compare the results of the chemical evolution
calculations for [Mn/Fe] of \emph{case $M_{Ch}$}, 
\emph{case $subM_{Ch}$}, and \emph{case mix} to observational data from the Galaxy.
In addition to the standard yields from \citet{woosley1995a} (which
trace the data along the lower edge at $[\mathrm{Fe}/\mathrm{H}] \lesssim -1.0$),
we also include evolution models with their Mn yield enhanced by 25 per cent (thick lines).
These Mn-enhanced models demonstrate that the final Mn at high metallicity
is rather insensitive to the assumed massive star yields at low metallicity.
Naturally, owing to the sub-solar production ratio of [Mn/Fe] of sub-\mch\ based 
SNe~Ia explosions, \emph{case $subM_{Ch}$} (blue lines) falls short of reproducing the observed trend.
The results of \emph{case $M_{Ch}$} (red lines) on the other hand reach and actually exceed 
the solar abundance. The data are best reproduced by a scenario where both sub-\mch\ and 
near-\mch\ primaries are present at roughly equal proportions (purple lines).
These results are a clear indication that SNe~Ia cannot exclusively
stem from sub-\mch\ WD primaries, due to their inability to produce 
enough Mn, as compared to the solar abundance.
In Fig.\ \ref{fig1} (bottom), we show the results of the chemical evolution
calculations for [Mn/Fe] of \emph{case $M_{Ch}+$} and \emph{case $subM_{Ch}$+2002cx}.
It is evident that using the metallicity dependent yields (red dashed)
reduces [Mn/Fe] somewhat, but the effect is of secondary nature.
In light of \citet{pakmor2013a}, we note that 
\emph{case $subM_{Ch}$+2002cx} (blue dashed) also falls significantly short of reaching solar [Mn/Fe],
even though \emph{case $subM_{Ch}$+2002cx} assumes a very high fraction of 
2002cx-like SNe (the expected relative fraction SN~2002cx-like SNe is around 4 per cent, \citealt{li2011c}).
Although model N5def has almost the same [Mn/Fe] production factor as the N100 model,
it produces much less Fe and Mn in total (a factor of ${\sim}3.5$ less, which is expeced to be
typical for the faint SN~2002cx-like objects), which explains its relatively small impact on [Mn/Fe].

\section{Conclusions}
\label{sec:conclusions}
The observed abundance trend of [Mn/Fe] at [Fe/H]
$\gtrsim 0.0$ suggests that sub-\mch\ WD primaries cannot be the
only progenitors producing SNe~Ia in the Galaxy;
either only near-\mch\ primary WDs or a combination of near-\mch\ and sub-\mch\ 
primaries (a mix of equal parts results in a good match to data) 
is needed to reach the observed [Mn/Fe] in the Sun.
\citet{mennekens2012a} reached a similar conclusion. They found that 
to reproduce the metallicity distribution of G-type stars in the solar
neighbourhood, both SDS and DDS progenitors must contribute to
the Galactic population of SNe~Ia.
Based on our chemical evolution calculations, we can also exclude 
that a combination of sub-\mch\ WD primaries and near-\mch\ WD primaries exploding 
as pure deflagrations that only partially unbind the primary (i.e. 2002cx-like SNe) 
constitute the entirety of SN~Ia progenitors. 

We speculate that the discrepancy between the chemical evolution of Mn in
dwarf spheroidal galaxies (dSph) and in the Milky Way \citep[see][]{mcwilliam2003a,north2012a}
could also be explained if SNe~Ia are not arising from a unique channel.
A different relative frequency of near-\mch\ and sub-\mch\ primaries 
(e.g.\ due to star formation history or metallicity)
could also be a solution to the Mn problem in dSph, since this would have 
an overall similar effect as the strong intrinsic dependency
on metallicity of the Mn yields invoked by \citet{cescutti2008a}.
In closing, we caution that any 
effect that raises [Mn/Fe] for sub-\mch\ primary explosion
models to super-solar would remove the need for a large fraction of near-\mch\ primaries.

\begin{acknowledgements}
  I.R.S.\ was funded by the Deutsche
  Forschungsgemein\-schaft (DFG) through the graduate school on
  ``Theoretical Astrophysics and Particle Physics'' (GRK 1147).
  F.~K.~R.\ was supported by the DFG
  via the Emmy Noether Program (RO 3676/1-1)
  and by the ARCHES prize of the German Federal Ministry of Education
  and Research (BMBF), and R.~P.\ by the European Research
  Council under ERC-StG grant EXAGAL-308037. Funding for collaboration was provided by the
  DAAD/Go8 German-Australian exchange program.
\end{acknowledgements}

\bibliographystyle{aa} \bibliography{astrofritz}

\end{document}